\documentclass[a4paper,11pt]{article}
\usepackage{pos}
\usepackage{fancyhdr}
\usepackage{xspace}
\usepackage{tabularx}
\usepackage{multirow}
\usepackage{booktabs}
\usepackage{amsmath}
\usepackage{subcaption}
\usepackage[capitalise]{cleveref}

\usepackage{wrapfig}
\usepackage{enumitem}
\usepackage[utf8]{inputenc}
\usepackage{array,multirow,graphicx}

\usepackage{lineno}

\def\Xmax{X_\text{max}}

\def\R{\widetilde{R}}

\def\gcm{\text{g/cm$^2$}}

\def\orcid#1{\href{https://orcid.org/#1}{\includegraphics[height=1.55ex]{orcid-ID}}}

\def\un2{\texttt{Universality\,II}}

\title{Update on the intermediate arrival-direction analyses of the Pierre Auger Observatory}
\ShortTitle{Update on the intermediate arrival-direction analyses of Auger}

\author*[ab]{Lorenzo Apollonio}

\affiliation[a]{Università degli Studi di Milano, Physics Department, via G. Celoria 16, Milan, Italy}
\affiliation[b]{Istituto Nazionale di Fisica Nucleare, Sezione di Milano}

\onbehalf{for the Pierre Auger Collaboration$^c$}
\affiliation[c]{Observatorio Pierre Auger, Av.\ San Mart{\'\i}n Norte 304, 5613 Malarg\"ue, Argentina\\
Full author list:\normalfont{\url{https://www.auger.org/archive/authors_icrc_2025.html}}}

\emailAdd{spokespersons@auger.org}

\abstract{
  We present an update on the arrival-direction analyses conducted on intermediate angular scales using the complete Phase I data of the Pierre Auger Observatory up to the end of $2022$ with a total exposure of $135{,}000\,\text{km}^2\,\text{sr}\,\text{yr}$. 
  We show the arrival-direction distribution of the ultra-high-energy cosmic rays along the supergalactic plane above $20\,\text{EeV}$,  and an update in the search for magnetically-induced signatures in the arrival directions. 
  Furthermore, we present the potential of introducing estimators for the rigidity ordering of the events to enhance arrival-direction analyses on small to intermediate angular scales. 
  To achieve this, we take advantage of two estimators working on the response of the surface detector: an analytical fit based on air-shower universality and a deep neural network.
}

\ConferenceLogo{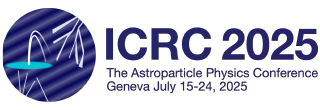}

\FullConference{39th International Cosmic Ray Conference (ICRC2025)\\
 15–24 July 2025\\
Geneva, Switzerland\\}
\begin{document}
\maketitle

\section{Introduction}

The Pierre Auger Observatory~\cite{ref:pao} is the largest observatory in the world studying ultra-uigh-energy cosmic rays (UHECRs).
The analyses completed on the arrival directions of the events are a powerful tool  to shed light about their astrophysical sources.
The Auger Collaboration has reported the presence of a dipolar modulation in right ascension for events above $8\,\text{EeV}$\footnote{$1\,\text{EeV}=10^{18}\,\text{eV}$}~\cite{ref:dipole}. 
The modulation has a significance of $6.8\sigma$, and its direction points $\sim115^\circ$ away from the Galactic Centre providing evidence that the anisotropy observed is of extragalactic origin. 
Furthermore, the Auger Collaboration has conducted analyses on smaller angular scales at energies above $32\,\text{EeV}$~\cite{ref:auger_icrc23_ad,ref:apj22}.
The most promising result has been the observation of an excess in the Centaurus region, with a post-trial significance of $4.0\sigma$.
This excess is also the driving force of an indication of correlation with a catalog of starburst galaxies, with a post-trial significance of $3.8\sigma$.

In this proceeding, we present an update on the arrival-direction analyses completed on intermediate angular scales with the data of the Pierre Auger Observatory.
The analyses include a scan looking for overdensities along the supergalactic plane and a search for magnetically-induced signatures.
The scan on the supergalactic includes events above $20\,\text{EeV}$ in an analysis on intermediate angular scales.

Moreover, we test on simulations the potential of introducing estimators for the rigidity ordering of the events, inferred with the data registered with the surface detector, in the analyses on intermediate angular scales.
The estimators are based on the indirect determination of the depth of shower maximum ($\Xmax$), using two possible methods: a fit based on air-shower universality \cite{ref:univ,ref:univ_icrc} and a deep neural network (DNN) \cite{ref:dnn}.
These methods have worse resolutions on the $\Xmax$ indirect determination  ($\approx40\,\text{g}\,\text{cm}^{-2}$ for the universality fit and $\approx 25\,\text{g}\,\text{cm}^{-2}$ for the DNN) respect to the fluorescence-detector measurement ($\approx15\,\text{g}\,\text{cm}^{-2}$).
However, since these estimators can be applied to the data of the surface detector, this ensures $\sim10$ times more events than with the fluorescence detector.

\section{The data set}

The data set used includes the events recorded with the water-Cherenkov surface detectors of the Pierre Auger Observatory from $1^\text{st}$ January 2004 to $31^\text{st}$ December 2022, namely the Phase~I of the Pierre Auger Observatory.
In the transition phase, during the deployment of the AugerPrime upgrade~\cite{Berat:2023ttm} (during 2021 and 2022), only events reconstructed with stations with pre-update electronics are included.
We consider events with a zenith angle $\theta<80^\circ$, equal to a field of view ranging from $-90^\circ$ to $+45^\circ$ in declination.
Two different selection criteria are applied for events with $\theta<60^\circ$ (vertical events) and events with $60^\circ\leq\theta<80^\circ$ (inclined events).
For vertical events, we require that the station with largest signal is surrounded by at least 4 active stations and that the reconstructed shower core is within an isosceles triangle of active stations.
For inclined events, we require that the station closest to the reconstructed shower core is surrounded by at least 5 active stations.
These criteria ensure an angular resolution $\lesssim 1^\circ$ for both vertical and inclined events, for the energies considered in this work.
The statistical uncertainty on the energy is $\sim7\%$ and the systematic uncertainty on the absolute energy scale is $14\%$ ~\cite{ref:prd2020, ref:jinst2020}.
The total exposure of this data set is $135{,}000\,\text{km}^2\,\text{sr}\,\text{yr}$.

\section{Supergalactic plane analysis}
\label{sec:sgp}

We start reporting the search for overdensities along the supergalactic plane (SGP) described in~\cite{ref:sgp}.
The analysis evaluates the probability of finding $N_\text{in}$ events inside a circular region of radius $\Psi=27^\circ$ when $N_\text{bg}$ events are expected.
The size of the circular region has been chosen according to the largest-significance overdensity region observed in the whole sky~\cite{ref:auger_icrc23_ad,ref:apj22}.
The centre positions of the regions are the bin centres of an HEALPix map with $N_\text{side}=64$ that meet the two following criteria: the SGP must be intersected by the circular region (i.e.\ the centre position has a supergalactic latitude $B$ with $-\Psi\leq B \leq+\Psi$) and the centre of the region must be contained in the field of view of the Observatory (declination $\delta\leq+45^\circ$).
We completed the analysis scanning over six energy thresholds, $E_\text{min}=20,25,32,40,50,63\,\text{EeV}$, i.e.\ $10^{19.3, 19.4, ...19.8}\,\text{eV}$ rounded to the nearest integer in EeV.

Given $N_\text{in}$ ($N_\text{out}$) events inside (outside) a circular region and evaluated the exposure $\varepsilon_\text{in}$ ($\varepsilon_\text{out}$) inside (outside) the region, we calculated the number of expected background events ($N_\text{bg}=N_\text{out}\varepsilon_\text{in}/\varepsilon_\text{out}$) and the ratio between the flux inside and outside $\phi_\text{in}/\phi_\text{out}=N_\text{in}/N_\text{out}$.

In \cref{tab:sgp_maxima}, we report the results obtained for the regions of the first and second maxima found along the SGP as a function of the selected energy threshold.
The requirement for the second maximum is that there is no overlap with the region of the first maximum.
For each maximum, we defined the frequentist $99\%$ confidence level upper limit to $\phi_\text{in}/\phi_\text{out}$ as the value of $\phi_\text{in}/\phi_\text{out}$ such that $\sum_{n=N_\text{in}+1}^{N_\text{tot}}P(n|N_\text{tot},\phi_\text{in}/\phi_\text{out})=0.01$.

Here, $P(n|N_\text{tot},\phi_\text{in}/\phi_\text{out})$ is the binomial probability of observing $n$ events in the region given $N_\text{tot}$ events.
In \cref{tab:sgp_maxima}, we show that the most significant excess is contained in the Centaurus region (for reference the position of the radiogalaxy Centaurus~A in supergalactic coordinates is $(L,B)=(159.7^\circ, -5.2^\circ)$), with an approximately stable position, throughout the energy-threshold range studied.
On the other hand, we show that the flux ratio increases, meaning that the particles of the excess have a different spectrum than the background ones, with a slower decrease with energy.

\begin{table}[]
    \begin{center}
	\resizebox{\textwidth}{!}{ 
    \begin{tabular}{cc|cccccccc|cccccccc}
         \multicolumn{2}{c}{} & \multicolumn{8}{c}{1st maximum} & \multicolumn{8}{c}{2nd maximun}\\
		 $E_\text{min}$ & $N_\text{tot}$ & $L$ & $B$ & $\frac{\varepsilon_\text{in}}{\varepsilon_\text{tot}}$ & $N_\text{bg}$ & $N_\text{in}$ & $\frac{\phi_\text{in}}{\phi_\text{out}}$ & $Z_\text{LM}$ & ${}^{99\%}_\text{U.L.}$ & $L$ & $B$ & $\frac{\varepsilon_\text{in}}{\varepsilon_\text{tot}}$ & $N_\text{bg}$ & $N_\text{in}$ & $\frac{\phi_\text{in}}{\phi_\text{out}}$ & $Z_\text{LM}$ & ${}^{99\%}_\text{U.L.}$ \\ 
         \midrule
		 $20\,\text{EeV}$ & 8832 & $162^\circ$ & $-6^\circ$ & $9.56\%$ & $829.$ & $990$ & $1.19^{+0.04}_{-0.04}$ & $+5.2\sigma$ & 1.29 &$241^\circ$ & $-5^\circ$ & $10.27\%$ & $900.$ & $971$ & $1.08^{+0.04}_{-0.04}$ & $+2.2\sigma$ & 1.17 \\ 
		 $25\,\text{EeV}$ & 5380 & $161^\circ$ & $-9^\circ$ & $9.56\%$ & $504.$ & $608$ & $1.21^{+0.05}_{-0.05}$ & $+4.2\sigma$ & 1.33 & $275^\circ$ & $-19^\circ$ & $8.00\%$ & $426.$ & $482$ & $1.13^{+0.05}_{-0.05}$ & $+2.6\sigma$ & 1.26 \\ 
		 $32\,\text{EeV}$ & 2936 & $163^\circ$ & $-8^\circ$ & $9.68\%$ & $276.$ & $363$ & $1.32^{+0.08}_{-0.07}$ & $+4.7\sigma$ & 1.50 & $276^\circ$ & $-17^\circ$ & $7.89\%$ & $229.$ & $264$ & $1.15^{+0.08}_{-0.07}$ & $+2.2\sigma$ & 1.34 \\ 
		 $40\,\text{EeV}$ & 1533 & $162^\circ$ & $-6^\circ$ & $9.56\%$ & $140.$ & $208$ & $1.49^{+0.11}_{-0.11}$ & $+5.1\sigma$ & 1.77 & $345^\circ$ & $-7^\circ$ & $1.00\%$ & $15.2.$ & $26$ & $1.71^{+0.36}_{-0.32}$ & $+2.5\sigma$ & 2.68 \\ 
		 $50\,\text{EeV}$ & 713 & $161^\circ$ & $-7^\circ$ & $9.56\%$ & $64.4$ & $103$ & $1.60^{+0.18}_{-0.16}$ & $+4.2\sigma$ & 2.05 & $322^\circ$ & $-22^\circ$ & $3.69\%$ & $25.9$ & $39$ & $1.51^{+0.26}_{-0.23}$ & $+2.4\sigma$ & 2.20 \\ 
		 $63\,\text{EeV}$ & 295 & $163^\circ$ & $-3^\circ$ & $9.56\%$ & $26.3.$ & $46$ & $1.75^{+0.30}_{-0.26}$ & $+3.3\sigma$ & 2.54 & $223^\circ$ & $+26^\circ$ & $9.56\%$ & $26.7$ & $42$ & $1.57^{+0.28}_{-0.25}$ & $+2.6\sigma$ & 2.31 \\ 
        \bottomrule
    \end{tabular}
    }
    \end{center}
    \caption{First and second maxima found in the SGP analysis, with the second maximum region not overlapping the first one. For each energy threshold, we report the number of events $N_\text{tot}$. For each maximum, we list the centre position in supergalactic coordinates $(L,B)$, the ratio of exposure inside and outside $\frac{\varepsilon_\text{in}}{\varepsilon_\text{tot}}$, the number of events inside ($N_\text{in}$) and outside ($N_\text{out}$), the flux ratio $\frac{\phi_\text{in}}{\phi_\text{out}}$, the Li-Ma significance ($Z_\text{LM}$) and the $99\%$ confidence level upper limit to $\phi_\text{in}/\phi_\text{out}$ (${}^{99\%}_\text{U.L.}$).}
    \label{tab:sgp_maxima}
\end{table}

We note that the Telescope Array Collaboration (TA) reported the presence of two overdensity regions intersecting the SGP.
These are an excess with $E\geq57\,\text{EeV}$ in a circular region centred in $(\alpha,\delta)\approx(145^\circ,+40^\circ)$ (a)~\cite{ref:ta_hotspot} and weaker excesses centred in $(\alpha,\delta)\approx(20^\circ,+45^\circ)$ with $E\geq10^{19.4,19.5,19.6}\,\text{eV}$ (b1, b2, b3)~\cite{ref:ta_excb,ref:ta_excb_icrc}.
As it is reported in \cref{tab:sgp_maxima}, the significance of the second maximum intersecting the SGP found for each energy threshold in our data is always $\sim2.5\sigma$.
The non observation of other excesses is in contrast with the results reported by the TA Collaboration, as the exposure of the Pierre Auger Observatory in the excess regions reported by the TA Collaboration is comparable with the TA one.
To better compare our results with those reported by the TA Collaboration, we changed our energy threshold according to the energy mismatch known between the two observatories (see Eq.\ (1) of~\cite{ref:pao_ta_en}) and completed the analysis for each excess region.
A full comparison of the results is reported in \cref{tab:ta_exc}.
We report that the significance obtained in each window is always $-0.7\sigma\lesssim Z_\text{LM} \lesssim +0.2\sigma$, in agreement with isotropy, while a significance of $\approx4\sigma$ is expected when considering the flux ratio observed by the TA Collaboration.
Based on the result reported by the TA Collaboration, we have been able to compute the frequentist $99\%$ lower limit on $\phi_\text{in}/\phi_\text{out}$ as the $\phi_\text{in}/\phi_\text{out}$ such that $\sum_{n=0}^{N_\text{in}-1}P(n|N_\text{tot},\phi_\text{in}/\phi_\text{out})=0.01$.
We note here that, while our result lays in agreement with isotropy, possible values of $\phi_\text{in}/\phi_\text{out}$ compatible with both the lower limits obtained from the results reported by the TA Collaboration and the upper limits obtained from our data, exist for all the reported TA overdensities.

\begin{table}[ht!]
    \begin{center}
	\resizebox{\textwidth}{!}{ 
    \begin{tabular}{c|ccccccccc|cccccccc}
         \multicolumn{1}{c}{} & \multicolumn{9}{c}{Telescope Array} & \multicolumn{8}{c}{Pierre Auger Observatory} \\
         & $E_\text{min}$ & $N_\text{tot}$ & $\frac{\varepsilon_\text{in}}{\varepsilon_\text{tot}}$ & $N_\text{bg}$ & $N_\text{in}$ & $\frac{\phi_\text{in}}{\phi_\text{out}}$ & $Z_\text{LM}$ & ${}^{99\%}_\text{L.L.}$ & ${}^\text{post-}_\text{trial}$ & $E_\text{min}$ & $N_\text{tot}$ & $\frac{\varepsilon_\text{in}}{\varepsilon_\text{tot}}$ & $N_\text{bg}$ & $N_\text{in}$ & $\frac{\phi_\text{in}}{\phi_\text{out}}$ & $Z_\text{LM}$ & ${}^{99\%}_\text{U.L.}$ \\ 
        \midrule
        (a) & $57\,\text{EeV}$ & 216 & $9.47\%$ & 18.0 & 44 & $2.44^{+0.44}_{-0.39}$ & $+4.8\sigma$ & 1.60 & $2.8\sigma$ & $44.6\,\text{EeV}$ & 1074 & $1.00\%$ & 10.7 & 9 & $0.84^{+0.31}_{-0.25}$ & $-0.5\sigma$ & 1.76 \\
        (b1) & $10^{19.4}\,\text{eV}$ & 1125 & $5.88\%$ & 64.0 & 101 & $1.58^{+0.16}_{-0.17}$ & $+4.1\sigma$ & 1.22 & $3.3\sigma$ & $20.5\,\text{EeV}$ & 8374 & $0.84\%$ & 70.1 & 65 & $0.93^{+0.12}_{-0.11}$ & $-0.6\sigma$ & 1.23 \\
        (b2) & $10^{19.6}\,\text{eV}$ & 728 & $5.87\%$ & 41.1 & 70 & $1.70^{+0.22}_{-0.20}$ & $+4.0\sigma$ & 1.25 & $3.2\sigma$ & $25.5\,\text{EeV}$ & 5156 & $0.84\%$ & 43.5 & 39 & $0.90^{+0.15}_{-0.14}$ & $-0.7\sigma$ & 1.29 \\
        (b3) & $10^{19.8}\,\text{eV}$ & 441 & $5.84\%$ & 24.6 & 45 & $1.83^{+0.31}_{-0.27}$ & $+3.6\sigma$ & 1.23 & $3.0\sigma$ & $31.7\,\text{EeV}$ & 2990 & $0.87\%$ & 26.0 & 27 & $1.04^{+0.21}_{-0.19}$ & $+0.2\sigma$ & 1.61 \\
        \midrule
    \end{tabular}
    }
    \end{center}
    \caption{Comparison between the result obtained by the Telescope Array Collaboration according to the latest update~\cite{ref:ta_excb_icrc} and the Auger Collaboration in the circular region of the excesses reported by Telescope Array Collaboration.}
    \label{tab:ta_exc}
\end{table}

\section{Search for magnetically-induced signatures}
We present here an update on the search of magnetically-induced signatures in UHECR arrival directions.
The deflection of a charged particle inside a regular magnetic field depends on the rigidity of the particle $R=E/Ze$ and the distance $L$ transverse within the magnetized regions.
If the deflections are small enough, the equation describing the observed arrival direction $\vec{\lambda}$ given the source direction $\vec{\lambda_s}$ can be linearized~\cite{Golup:2009zg} as $\vec{\lambda}=\vec{\lambda}_s+\frac{1}{E/Ze}\int_{0}^{L}d\vec{l}\times\vec{B}\simeq\vec{\lambda}_s+\frac{\vec{D}(\vec{\lambda}_s)}{E}$, where $\vec{D}(\vec{\lambda}_s)$ denotes the deflection power.
For $\mu \text{G}$ values typical of the Galactic magnetic field, we expect departures from the linear approximation below $\sim20\,\text{EV}$, thus for proton the approximation is valid above $\sim20\,\text{EeV}$, while for helium above $\sim40\,\text{EeV}$.
The signatures of this deflection would be the presence of sets of events (multiplets) oriented in the source direction within $\sim 20^\circ$ according to $1/E$.

We studied the presence of multiplets inside the Auger Phase~I data set with a high correlation coefficient between $\vec{\lambda}$ and $1/E$ using the same method as \cite{ref:mult}.
We decided to limit the data set to events above $40\,\text{EeV}$.
The correlation between a given set of $N$ events is calculated finding the projection $(x_i,y_i)$ of each event along the plane $(x,y)$, defined as the plane tangent to the celestial sphere in the average direction of the set of events.
We then compute the covariances $\text{Cov}(x,1/E)=\frac{1}{N}\sum_{i=1}^{N}(x_i-\langle x \rangle)(1/E_i-\langle 1/E \rangle)$ and $\text{Cov}(y,1/E)$.
We thus define the rotated plane $(u,w)$ as the plane where $\text{Cov}(u,1/E)$ is maximum and $\text{Cov}(w,1/E)=0$.
This plane is obtained by rotating $(x,y)$ with the rotation angle $\alpha=\arctan\bigl(\text{Cov}(y,1/E)/\text{Cov}(x,1/E)\bigr)$ between $u$ and $x$.
The correlation between $u$ and $1/E$ is thus given by $C(u,1/E)=\frac{\text{Cov}(u,1/E)}{\sqrt{\text{Var}(u)\text{Var}(1/E)}}$, where $\text{Var}(x)=\big\langle (x-\langle x \rangle)^2 \big\rangle$.

To identify a set of $N$ events as a multiplet, we require three conditions: first, the correlation coefficient has to be sufficiently large ($C(u,1/E)>C_\text{min}$), second, the set of events does not spread largely across the $w$-axis ($\max(w_i)<w_\text{max}$), and finally the events do not spread largely across the sky ($d_{ij}<20^\circ$, where $d_{ij}$ is the angular distance between the $i$-th and $j$-th events of the set).
The values of $C_\text{min}$ and $w_\text{max}$, chosen to maximize the signal enhancement and minimize the isotropic background chance probability, are $C_\text{min}=0.85$ and $w_\text{max}=1.5^\circ$.

We completed an all-sky analysis, where no condition on the multiplet position is defined, and a targeted analysis, where multiplets are searched around specific sources.
In the targeted analysis, we imposed that the position of the reconstructed source is within $3^\circ$ from the selected source.
We performed the targeted analysis considering ten potential sources as in \cite{ref:mult}, among which, three are AGNs (Centaurus~A, M~87 and Fornax~A) and seven are SBGs (NGC~253, NGC~4945, Circinus, M~83, NGC~4631, NGC~1808, NGC~1068).
The $p$-value of $n$ independent multiplets with multiplicity equal or larger than $m$ is given by the fraction of isotropic simulations where at least $n$ or more multiplets are found with multiplicity equal or larger than $m$, passing the same cuts.

For the all-sky analysis, the largest multiplet we have obtained is a 13-plet with an isotropic chance probability of $12\%$.
For the targeted analysis, the largest multiplet is a 9-plet for Centaurus~A, with a chance probability of $6\%$.

\section{Rigidity-ordering estimators for arrival directions}
\label{sec:mead}

Given that magnetic deflections scale with the rigidity of the particles, events with higher rigidity are expected to show more pronounced anisotropies.
We conducted a series of simulations to determine the most effective strategy to enhance a potential excess of high-rigidity events in the Centaurus region, defined as the circular area within $27^\circ$ of the radiogalaxy Centaurus~A (Cen\,A)\cite{ref:apj22,ref:auger_icrc23_ad}, using rigidity-ordering estimators applied to the data registered with the surface detector.
Assuming that the observed excess is not due to statistical fluctuations, our goal is to study how discarding low-rigidity events can maximize the statistical significance of the excess, under different hypothesis for the mass composition of the events that are located in the overdensity region.
The analysis was restricted to events with reconstructed energy above $38\,\text{EeV}$, which is the energy threshold where the maximum significance is found.

In each simulation, we defined a signal population $S$ confined to the Centaurus region, and a background population $B$ extending throughout the sky.
Since the depth of shower maximum ($\Xmax$) inferred with data from the surface detector can only be reconstructed for vertical events as of now, we randomly marked $N_\text{ver}^\text{sig}$ vertical events within the Centaurus region as signal, assigning them a mass according to the specific scenario under study.
All remaining vertical events were treated as background and assigned a mass according to the background model.

The events were then matched with air shower simulations of the corresponding mass, whose $\Xmax$ values had been reconstructed using both the universality-based method~\cite{ref:univ,ref:univ_icrc} and a DNN approach~\cite{ref:dnn}.
All simulations were performed using the EPOS-LHC hadronic interaction model~\cite{ref:epos-lhc}.

For each scenario, we tested three different values for the signal fraction in the Centaurus region ($f_\text{sig}^\text{Cen\,A} = 0.25$, $0.50$, and $0.75$), and performed 1000 independent sky simulations per case.
In this proceeding, we report the results for the following two scenarios.
\begin{enumerate}[label=(\roman*)]
    \item \textbf{helium scenario}: the signal composition is made of helium particles, while the background composition follows the "Auger mix" composition.
    This composition is obtained fitting the $\Xmax$ distributions~\cite{ref:olena}, measured using the fluorescence detector, with the expectation based on the EPOS-LHC hadronic interaction model~\cite{ref:epos-lhc}, asssuming four atomic species: proton, helium, nitrogen and iron;
    \item \textbf{mixed composition scenario}: in this scenario, we account for possible deflections and place lower-rigidity events further away from Centaurus~A.
    We considered that the cosmic rays are deflected due to turbulent extragalactic magnetic fields following a von Mises distribution ($p_\text{VM}(d, R)$), where $d$ is the angular distance from Centaurus~A and $R$ is the rigidity of the event.
    The deflection for an event with atomic number $Z_i$ and energy $E$ is defined by the concentration parameter $k_i= \frac{1}{\Theta_0^2}\bigl(\frac{E/Z_ie}{10\,\text{EV}}\bigr)$.
    We impose $\Theta_0=16.8^\circ$. 
    This value was taken from the outcome of the scenario considering Centaurus~A as UHECR source in \cite{ref:combined_fit_ad}.
    The probability for an event distant $d$ from Centaurus~A, to be of the atomic species $Z_i$ is given by $w_i=p_\text{VM}(d,E/Z_ie)/\sum_jp_\text{VM}(d,E/Z_je)$.
    We consider a signal mixed population of proton, helium and oxygen, i.e.\ $Z_i\in[1,2,8]$.
    As in the previous scenario, the background population follows the "Auger mix" composition model.
\end{enumerate}

Starting from the simulated $\Xmax$, we subtracted the decadal elongation rate, defining the energy-independent mass-sensitive parameter $X_\text{max}^{19}=X_\text{max}-D\lg(E/10^{19}\,\text{eV})$.
We set $D=58\,\text{g/cm}^2$, which is approximately equal in all hadronic interaction models.
We then introduced the rigidity proxy parameter $\widetilde{R} := E_0 / \widetilde{Z}e$, using $1/\widetilde{Z} := 2 \; \exp{\left((\Xmax^{19}-x_\text{ref})/\lambda\right)}$.
In the previous equation, $\lambda\simeq 22.3\,\gcm$ represents the increase in the average depth of the shower maximum per unit step in $\ln(A)$ based on the Heitler-Matthews model, and is approximately invariant in all hadronic interaction models.
$\widetilde{R}$ is not the absolute rigidity of the events, but approximately preserves the rigidity ordering of the data set, given an ideal reconstruction.
Thus, we ordered the simulated data sets according to $\R$, and discarded events as a function of increasing $\R$ to study the evolution with $\R$ of the significance in the Centaurus~A targeted analysis.
We note that the ordering is independent of the choice of $x_\text{ref}$.

\begin{figure}
    \centering
    \begin{subfigure}{0.85\linewidth}  
        \centering
        \includegraphics[width=\linewidth]{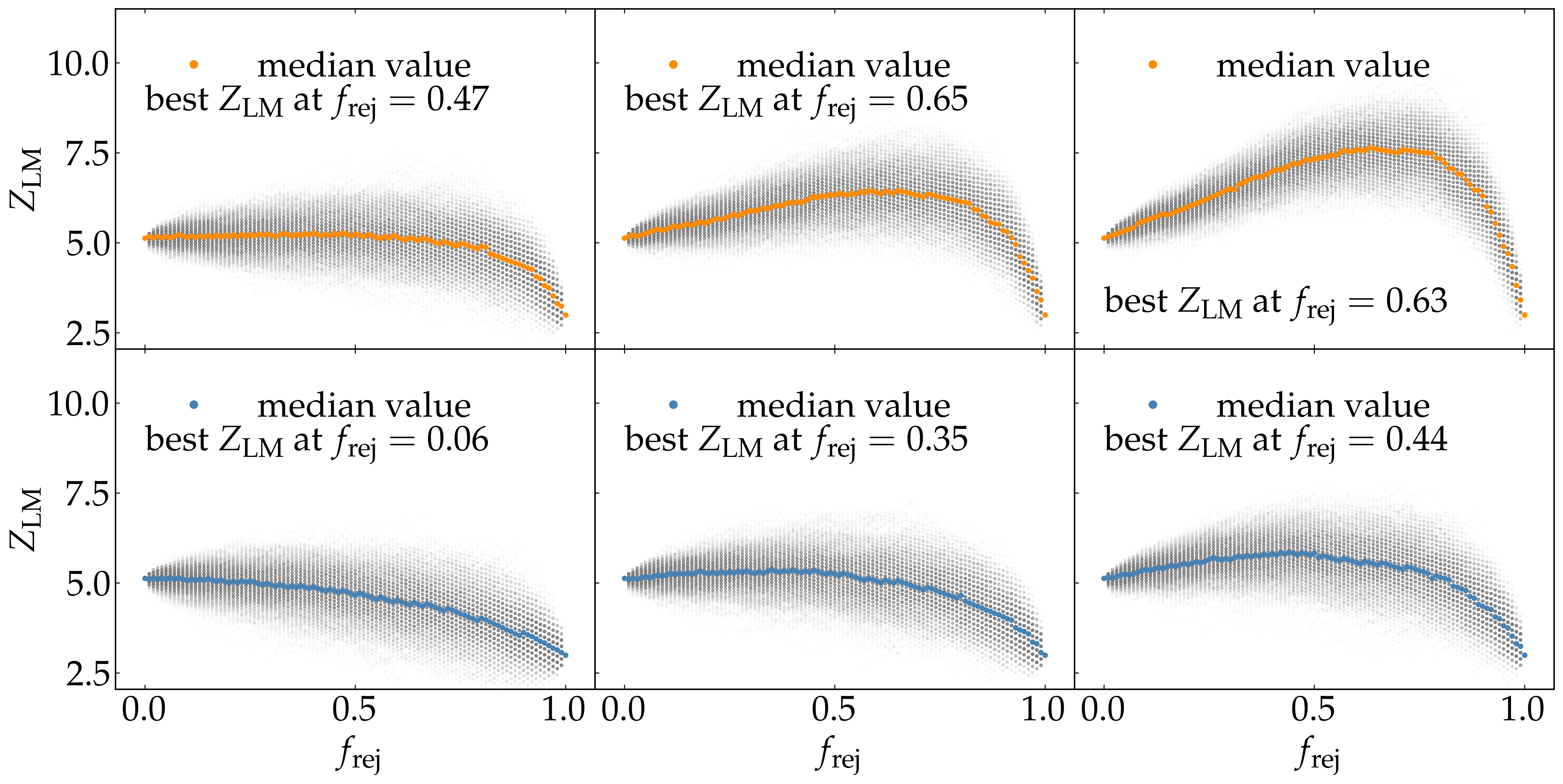}
        \caption{Universality}  
    \end{subfigure}
    \begin{subfigure}{0.85\linewidth}
        \centering
        \includegraphics[width=\linewidth]{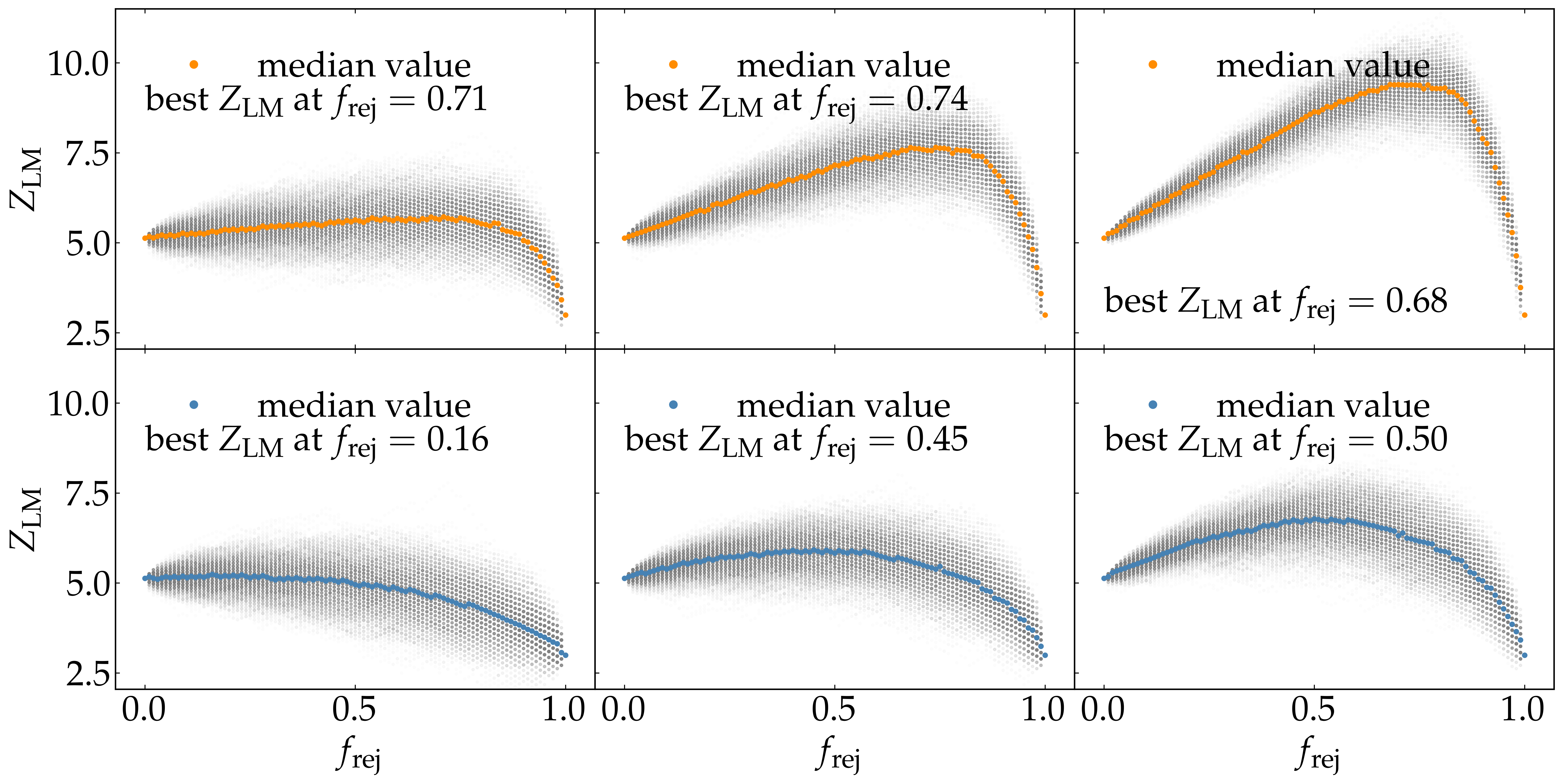}
        \caption{DNN}  
    \end{subfigure}
    \caption{Evolution of the Li-Ma significance ($Z_\text{LM}$) of the Centaurus region excess as a function of the fraction of the low-rigidity events rejected ($f_\text{rej}$) , considering the estimator based on Universality (a) and the one inferred with DNN (b). In each panel, the top row corresponds to the helium scenario while the bottom one corresponds to the mixed composition case. The left, middle and right columns are the $f_\text{sig}^\text{Cen\,A}=$0.25, 0.5 and 0.75 cases, respectively.}
    \label{fig:cena_evol}
\end{figure}

In \cref{fig:cena_evol}, we present the Li-Ma significance ($Z_\text{LM}$) of the Centaurus A excess as a function of the fraction of low-rigidity events rejected ($f_\text{rej}$). 
The top panel corresponds to the results when the estimator based on air-shower universality is used, and the bottom panel corresponds to the results when the estimator inferred with the DNN is considered.
In each panel, the top and bottom rows correspond to the helium and mix composition scenario, while the left, middle and right columns are the $f_\text{sig}^\text{Cen\,A}=$0.25, 0.5, and 0.75 cases, respectively.
With this figure we illustrate that the introduction of rigidity-ordering estimators can significantly enhance the significance of an excess caused by high-rigidity events around Centaurus~A\@. 
As expected, the enhancement is more pronounced when there is a larger mass difference between signal and background, and it increases with the signal fraction. 
Moreover, the increase is larger when the estimator inferred with DNN is considered, given that it has a better resolution than the universality-based one.
For both methods, the maximum significance is obtained when a substantial fraction of low-rigidity events is discarded ($f_\text{rej} \sim 0.4-0.6$).
We note, however, that these results are obtained in the specific assumption that a high-rigidity signal is present only in the Centaurus~A region; thus results on data may vary significantly.

\section{Conclusion}
We presented here an update on the arrival-direction analyses on intermediate angular scale with data of the Phase~I of the Pierre Auger Observatory.

Firstly, we showed a blind search looking for overdensities on circular regions along the SGP with radius fixed to $27^\circ$.
The most significant overdensity is located in the Centaurus region with a position approximately stable at all energy thresholds considered.
The largest significance is found above $20\,\text{EeV}$ where $Z_\text{LM}=+5.2\sigma$.
We also studied the regions along the SGP where the TA Collaboration has reported overdensities in their data.
With comparable statistics, all the Li-Ma significances obtained are in agreement with isotropy, while significances of $\approx4\sigma$ were expected. 
We note however that there are values of the flux ratio compatible with the upper limits obtained from our data and the lower limits inferred from the results reported by the TA Collaboration.

We then reported an update on the search for multiplets.
By completing an all-sky analysis, the largest multiplet observed is one $13$-plet with a probability of $12\%$ of being present in an isotropic distribution of arrival directions.
By doing a targeted analysis looking for multiplets within $3^\circ$ from selected sources, we observed one $9$-plet near Centaurus~A with a chance probability of $6\%$.

Lastly,  using simulations, we studied the introduction of estimators for the rigidity ordering for the data recorded with the surface detector in the Centaurus~A targeted analysis.
We showed that if the excess observed in the Centaurus region is formed by high-rigidity events, we expect to improve the significance of the Centaurus~A targeted analysis by discarding a moderate fraction of low-rigidity events.

\clearpage
\section*{The Pierre Auger Collaboration}

{\footnotesize\setlength{\baselineskip}{10pt}
\noindent
\begin{wrapfigure}[11]{l}{0.12\linewidth}
\vspace{-4pt}
\includegraphics[width=0.98\linewidth]{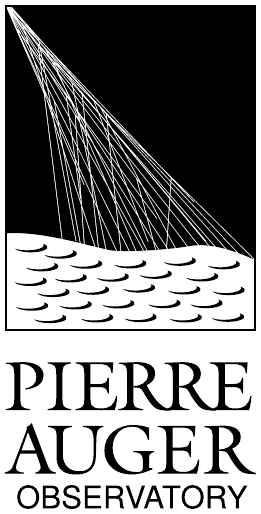}
\end{wrapfigure}
\begin{sloppypar}\noindent
% created on 2025-06-06
A.~Abdul Halim$^{13}$,
P.~Abreu$^{70}$,
M.~Aglietta$^{53,51}$,
I.~Allekotte$^{1}$,
K.~Almeida Cheminant$^{78,77}$,
A.~Almela$^{7,12}$,
R.~Aloisio$^{44,45}$,
J.~Alvarez-Mu\~niz$^{76}$,
A.~Ambrosone$^{44}$,
J.~Ammerman Yebra$^{76}$
% ...

\end{sloppypar}
\begin{center}
\end{center}

\vspace{1ex}
% created on 2025-06-06
% needs \usepackage{enumitem}
\begin{description}[labelsep=0.2em,align=right,labelwidth=0.7em,labelindent=0em,leftmargin=2em,noitemsep,before={\renewcommand\makelabel[1]{##1 }}]
\item[$^{1}$] Centro At\'omico Bariloche and Instituto Balseiro (CNEA-UNCuyo-CONICET), San Carlos de Bariloche, Argentina
\item[$^{2}$] Departamento de F\'\i{}sica and Departamento de Ciencias de la Atm\'osfera y los Oc\'eanos, FCEyN, Universidad de Buenos Aires and CONICET, Buenos Aires, Argentina
\item[$^{3}$] IFLP, Universidad Nacional de La Plata and CONICET, La Plata, Argentina
\item[$^{4}$] Instituto de Astronom\'\i{}a y F\'\i{}sica del Espacio (IAFE, CONICET-UBA), Buenos Aires, Argentina
\item[$^{5}$] Instituto de F\'\i{}sica de Rosario (IFIR) -- CONICET/U.N.R.\ and Facultad de Ciencias Bioqu\'\i{}micas y Farmac\'euticas U.N.R., Rosario, Argentina
\item[$^{6}$] Instituto de Tecnolog\'\i{}as en Detecci\'on y Astropart\'\i{}culas (CNEA, CONICET, UNSAM), and Universidad Tecnol\'ogica Nacional -- Facultad Regional Mendoza (CONICET/CNEA), Mendoza, Argentina
\item[$^{7}$] Instituto de Tecnolog\'\i{}as en Detecci\'on y Astropart\'\i{}culas (CNEA, CONICET, UNSAM), Buenos Aires, Argentina
\item[$^{8}$] International Center of Advanced Studies and Instituto de Ciencias F\'\i{}sicas, ECyT-UNSAM and CONICET, Campus Miguelete -- San Mart\'\i{}n, Buenos Aires, Argentina
\item[$^{9}$] Laboratorio Atm\'osfera -- Departamento de Investigaciones en L\'aseres y sus Aplicaciones -- UNIDEF (CITEDEF-CONICET), Argentina
\item[$^{10}$] Observatorio Pierre Auger, Malarg\"ue, Argentina
\item[$^{11}$] Observatorio Pierre Auger and Comisi\'on Nacional de Energ\'\i{}a At\'omica, Malarg\"ue, Argentina
\item[$^{12}$] Universidad Tecnol\'ogica Nacional -- Facultad Regional Buenos Aires, Buenos Aires, Argentina
\item[$^{13}$] University of Adelaide, Adelaide, S.A., Australia
\item[$^{14}$] Universit\'e Libre de Bruxelles (ULB), Brussels, Belgium
\item[$^{15}$] Vrije Universiteit Brussels, Brussels, Belgium
\item[$^{16}$] Centro Brasileiro de Pesquisas Fisicas, Rio de Janeiro, RJ, Brazil
\item[$^{17}$] Centro Federal de Educa\c{c}\~ao Tecnol\'ogica Celso Suckow da Fonseca, Petropolis, Brazil
\item[$^{18}$] Instituto Federal de Educa\c{c}\~ao, Ci\^encia e Tecnologia do Rio de Janeiro (IFRJ), Brazil
\item[$^{19}$] Universidade de S\~ao Paulo, Escola de Engenharia de Lorena, Lorena, SP, Brazil
\item[$^{20}$] Universidade de S\~ao Paulo, Instituto de F\'\i{}sica de S\~ao Carlos, S\~ao Carlos, SP, Brazil
\item[$^{21}$] Universidade de S\~ao Paulo, Instituto de F\'\i{}sica, S\~ao Paulo, SP, Brazil
\item[$^{22}$] Universidade Estadual de Campinas (UNICAMP), IFGW, Campinas, SP, Brazil
\item[$^{23}$] Universidade Estadual de Feira de Santana, Feira de Santana, Brazil
\item[$^{24}$] Universidade Federal de Campina Grande, Centro de Ciencias e Tecnologia, Campina Grande, Brazil
\item[$^{25}$] Universidade Federal do ABC, Santo Andr\'e, SP, Brazil
\item[$^{26}$] Universidade Federal do Paran\'a, Setor Palotina, Palotina, Brazil
\item[$^{27}$] Universidade Federal do Rio de Janeiro, Instituto de F\'\i{}sica, Rio de Janeiro, RJ, Brazil
\item[$^{28}$] Universidad de Medell\'\i{}n, Medell\'\i{}n, Colombia
\item[$^{29}$] Universidad Industrial de Santander, Bucaramanga, Colombia
\item[$^{30}$] Charles University, Faculty of Mathematics and Physics, Institute of Particle and Nuclear Physics, Prague, Czech Republic
\item[$^{31}$] Institute of Physics of the Czech Academy of Sciences, Prague, Czech Republic
\item[$^{32}$] Palacky University, Olomouc, Czech Republic
\item[$^{33}$] CNRS/IN2P3, IJCLab, Universit\'e Paris-Saclay, Orsay, France
\item[$^{34}$] Laboratoire de Physique Nucl\'eaire et de Hautes Energies (LPNHE), Sorbonne Universit\'e, Universit\'e de Paris, CNRS-IN2P3, Paris, France
\item[$^{35}$] Univ.\ Grenoble Alpes, CNRS, Grenoble Institute of Engineering Univ.\ Grenoble Alpes, LPSC-IN2P3, 38000 Grenoble, France
\item[$^{36}$] Universit\'e Paris-Saclay, CNRS/IN2P3, IJCLab, Orsay, France
\item[$^{37}$] Bergische Universit\"at Wuppertal, Department of Physics, Wuppertal, Germany
\item[$^{38}$] Karlsruhe Institute of Technology (KIT), Institute for Experimental Particle Physics, Karlsruhe, Germany
\item[$^{39}$] Karlsruhe Institute of Technology (KIT), Institut f\"ur Prozessdatenverarbeitung und Elektronik, Karlsruhe, Germany
\item[$^{40}$] Karlsruhe Institute of Technology (KIT), Institute for Astroparticle Physics, Karlsruhe, Germany
\item[$^{41}$] RWTH Aachen University, III.\ Physikalisches Institut A, Aachen, Germany
\item[$^{42}$] Universit\"at Hamburg, II.\ Institut f\"ur Theoretische Physik, Hamburg, Germany
\item[$^{43}$] Universit\"at Siegen, Department Physik -- Experimentelle Teilchenphysik, Siegen, Germany
\item[$^{44}$] Gran Sasso Science Institute, L'Aquila, Italy
\item[$^{45}$] INFN Laboratori Nazionali del Gran Sasso, Assergi (L'Aquila), Italy
\item[$^{46}$] INFN, Sezione di Catania, Catania, Italy
\item[$^{47}$] INFN, Sezione di Lecce, Lecce, Italy
\item[$^{48}$] INFN, Sezione di Milano, Milano, Italy
\item[$^{49}$] INFN, Sezione di Napoli, Napoli, Italy
\item[$^{50}$] INFN, Sezione di Roma ``Tor Vergata'', Roma, Italy
\item[$^{51}$] INFN, Sezione di Torino, Torino, Italy
\item[$^{52}$] Istituto di Astrofisica Spaziale e Fisica Cosmica di Palermo (INAF), Palermo, Italy
\item[$^{53}$] Osservatorio Astrofisico di Torino (INAF), Torino, Italy
\item[$^{54}$] Politecnico di Milano, Dipartimento di Scienze e Tecnologie Aerospaziali , Milano, Italy
\item[$^{55}$] Universit\`a del Salento, Dipartimento di Matematica e Fisica ``E.\ De Giorgi'', Lecce, Italy
\item[$^{56}$] Universit\`a dell'Aquila, Dipartimento di Scienze Fisiche e Chimiche, L'Aquila, Italy
\item[$^{57}$] Universit\`a di Catania, Dipartimento di Fisica e Astronomia ``Ettore Majorana``, Catania, Italy
\item[$^{58}$] Universit\`a di Milano, Dipartimento di Fisica, Milano, Italy
\item[$^{59}$] Universit\`a di Napoli ``Federico II'', Dipartimento di Fisica ``Ettore Pancini'', Napoli, Italy
\item[$^{60}$] Universit\`a di Palermo, Dipartimento di Fisica e Chimica ''E.\ Segr\`e'', Palermo, Italy
\item[$^{61}$] Universit\`a di Roma ``Tor Vergata'', Dipartimento di Fisica, Roma, Italy
\item[$^{62}$] Universit\`a Torino, Dipartimento di Fisica, Torino, Italy
\item[$^{63}$] Benem\'erita Universidad Aut\'onoma de Puebla, Puebla, M\'exico
\item[$^{64}$] Unidad Profesional Interdisciplinaria en Ingenier\'\i{}a y Tecnolog\'\i{}as Avanzadas del Instituto Polit\'ecnico Nacional (UPIITA-IPN), M\'exico, D.F., M\'exico
\item[$^{65}$] Universidad Aut\'onoma de Chiapas, Tuxtla Guti\'errez, Chiapas, M\'exico
\item[$^{66}$] Universidad Michoacana de San Nicol\'as de Hidalgo, Morelia, Michoac\'an, M\'exico
\item[$^{67}$] Universidad Nacional Aut\'onoma de M\'exico, M\'exico, D.F., M\'exico
\item[$^{68}$] Institute of Nuclear Physics PAN, Krakow, Poland
\item[$^{69}$] University of \L{}\'od\'z, Faculty of High-Energy Astrophysics,\L{}\'od\'z, Poland
\item[$^{70}$] Laborat\'orio de Instrumenta\c{c}\~ao e F\'\i{}sica Experimental de Part\'\i{}culas -- LIP and Instituto Superior T\'ecnico -- IST, Universidade de Lisboa -- UL, Lisboa, Portugal
\item[$^{71}$] ``Horia Hulubei'' National Institute for Physics and Nuclear Engineering, Bucharest-Magurele, Romania
\item[$^{72}$] Institute of Space Science, Bucharest-Magurele, Romania
\item[$^{73}$] Center for Astrophysics and Cosmology (CAC), University of Nova Gorica, Nova Gorica, Slovenia
\item[$^{74}$] Experimental Particle Physics Department, J.\ Stefan Institute, Ljubljana, Slovenia
\item[$^{75}$] Universidad de Granada and C.A.F.P.E., Granada, Spain
\item[$^{76}$] Instituto Galego de F\'\i{}sica de Altas Enerx\'\i{}as (IGFAE), Universidade de Santiago de Compostela, Santiago de Compostela, Spain
\item[$^{77}$] IMAPP, Radboud University Nijmegen, Nijmegen, The Netherlands
\item[$^{78}$] Nationaal Instituut voor Kernfysica en Hoge Energie Fysica (NIKHEF), Science Park, Amsterdam, The Netherlands
\item[$^{79}$] Stichting Astronomisch Onderzoek in Nederland (ASTRON), Dwingeloo, The Netherlands
\item[$^{80}$] Universiteit van Amsterdam, Faculty of Science, Amsterdam, The Netherlands
\item[$^{81}$] Case Western Reserve University, Cleveland, OH, USA
\item[$^{82}$] Colorado School of Mines, Golden, CO, USA
\item[$^{83}$] Department of Physics and Astronomy, Lehman College, City University of New York, Bronx, NY, USA
\item[$^{84}$] Michigan Technological University, Houghton, MI, USA
\item[$^{85}$] New York University, New York, NY, USA
\item[$^{86}$] University of Chicago, Enrico Fermi Institute, Chicago, IL, USA
\item[$^{87}$] University of Delaware, Department of Physics and Astronomy, Bartol Research Institute, Newark, DE, USA
\item[] -----
\item[$^{a}$] Max-Planck-Institut f\"ur Radioastronomie, Bonn, Germany
\item[$^{b}$] also at Kapteyn Institute, University of Groningen, Groningen, The Netherlands
\item[$^{c}$] School of Physics and Astronomy, University of Leeds, Leeds, United Kingdom
\item[$^{d}$] Fermi National Accelerator Laboratory, Fermilab, Batavia, IL, USA
\item[$^{e}$] Pennsylvania State University, University Park, PA, USA
\item[$^{f}$] Colorado State University, Fort Collins, CO, USA
\item[$^{g}$] Louisiana State University, Baton Rouge, LA, USA
\item[$^{h}$] now at Graduate School of Science, Osaka Metropolitan University, Osaka, Japan
\item[$^{i}$] Institut universitaire de France (IUF), France
\item[$^{j}$] now at Technische Universit\"at Dortmund and Ruhr-Universit\"at Bochum, Dortmund and Bochum, Germany
\end{description}

% created on 2025-06-06
\section*{Acknowledgments}

\begin{sloppypar}
The successful installation, commissioning, and operation of the Pierre
Auger Observatory would not have been possible without the strong
commitment and effort from the technical and administrative staff in
Malarg\"ue. We are very grateful to the following agencies and
organizations for financial support:
\end{sloppypar}

\begin{sloppypar}
Argentina -- Comisi\'on Nacional de Energ\'\i{}a At\'omica; Agencia Nacional de
Promoci\'on Cient\'\i{}fica y Tecnol\'ogica (ANPCyT); Consejo Nacional de
Investigaciones Cient\'\i{}ficas y T\'ecnicas (CONICET); Gobierno de la
Provincia de Mendoza; Municipalidad de Malarg\"ue; NDM Holdings and Valle
Las Le\~nas; in gratitude for their continuing cooperation over land
access; Australia -- the Australian Research Council; Belgium -- Fonds
de la Recherche Scientifique (FNRS); Research Foundation Flanders (FWO),
Marie Curie Action of the European Union Grant No.~101107047; Brazil --
Conselho Nacional de Desenvolvimento Cient\'\i{}fico e Tecnol\'ogico (CNPq);
Financiadora de Estudos e Projetos (FINEP); Funda\c{c}\~ao de Amparo \`a
Pesquisa do Estado de Rio de Janeiro (FAPERJ); S\~ao Paulo Research
Foundation (FAPESP) Grants No.~2019/10151-2, No.~2010/07359-6 and
No.~1999/05404-3; Minist\'erio da Ci\^encia, Tecnologia, Inova\c{c}\~oes e
Comunica\c{c}\~oes (MCTIC); Czech Republic -- GACR 24-13049S, CAS LQ100102401,
MEYS LM2023032, CZ.02.1.01/0.0/0.0/16{\textunderscore}013/0001402,
CZ.02.1.01/0.0/0.0/18{\textunderscore}046/0016010 and
CZ.02.1.01/0.0/0.0/17{\textunderscore}049/0008422 and CZ.02.01.01/00/22{\textunderscore}008/0004632;
France -- Centre de Calcul IN2P3/CNRS; Centre National de la Recherche
Scientifique (CNRS); Conseil R\'egional Ile-de-France; D\'epartement
Physique Nucl\'eaire et Corpusculaire (PNC-IN2P3/CNRS); D\'epartement
Sciences de l'Univers (SDU-INSU/CNRS); Institut Lagrange de Paris (ILP)
Grant No.~LABEX ANR-10-LABX-63 within the Investissements d'Avenir
Programme Grant No.~ANR-11-IDEX-0004-02; Germany -- Bundesministerium
f\"ur Bildung und Forschung (BMBF); Deutsche Forschungsgemeinschaft (DFG);
Finanzministerium Baden-W\"urttemberg; Helmholtz Alliance for
Astroparticle Physics (HAP); Helmholtz-Gemeinschaft Deutscher
Forschungszentren (HGF); Ministerium f\"ur Kultur und Wissenschaft des
Landes Nordrhein-Westfalen; Ministerium f\"ur Wissenschaft, Forschung und
Kunst des Landes Baden-W\"urttemberg; Italy -- Istituto Nazionale di
Fisica Nucleare (INFN); Istituto Nazionale di Astrofisica (INAF);
Ministero dell'Universit\`a e della Ricerca (MUR); CETEMPS Center of
Excellence; Ministero degli Affari Esteri (MAE), ICSC Centro Nazionale
di Ricerca in High Performance Computing, Big Data and Quantum
Computing, funded by European Union NextGenerationEU, reference code
CN{\textunderscore}00000013; M\'exico -- Consejo Nacional de Ciencia y Tecnolog\'\i{}a
(CONACYT) No.~167733; Universidad Nacional Aut\'onoma de M\'exico (UNAM);
PAPIIT DGAPA-UNAM; The Netherlands -- Ministry of Education, Culture and
Science; Netherlands Organisation for Scientific Research (NWO); Dutch
national e-infrastructure with the support of SURF Cooperative; Poland
-- Ministry of Education and Science, grants No.~DIR/WK/2018/11 and
2022/WK/12; National Science Centre, grants No.~2016/22/M/ST9/00198,
2016/23/B/ST9/01635, 2020/39/B/ST9/01398, and 2022/45/B/ST9/02163;
Portugal -- Portuguese national funds and FEDER funds within Programa
Operacional Factores de Competitividade through Funda\c{c}\~ao para a Ci\^encia
e a Tecnologia (COMPETE); Romania -- Ministry of Research, Innovation
and Digitization, CNCS-UEFISCDI, contract no.~30N/2023 under Romanian
National Core Program LAPLAS VII, grant no.~PN 23 21 01 02 and project
number PN-III-P1-1.1-TE-2021-0924/TE57/2022, within PNCDI III; Slovenia
-- Slovenian Research Agency, grants P1-0031, P1-0385, I0-0033, N1-0111;
Spain -- Ministerio de Ciencia e Innovaci\'on/Agencia Estatal de
Investigaci\'on (PID2019-105544GB-I00, PID2022-140510NB-I00 and
RYC2019-027017-I), Xunta de Galicia (CIGUS Network of Research Centers,
Consolidaci\'on 2021 GRC GI-2033, ED431C-2021/22 and ED431F-2022/15),
Junta de Andaluc\'\i{}a (SOMM17/6104/UGR and P18-FR-4314), and the European
Union (Marie Sklodowska-Curie 101065027 and ERDF); USA -- Department of
Energy, Contracts No.~DE-AC02-07CH11359, No.~DE-FR02-04ER41300,
No.~DE-FG02-99ER41107 and No.~DE-SC0011689; National Science Foundation,
Grant No.~0450696, and NSF-2013199; The Grainger Foundation; Marie
Curie-IRSES/EPLANET; European Particle Physics Latin American Network;
and UNESCO.
\end{sloppypar}

}

\end{document}